\documentclass [a4paper] {article}

\usepackage[english]{babel}

\usepackage[margin=1.5cm]{geometry}
\usepackage{amsmath}
\usepackage{amsfonts}

\usepackage{pgfplots}
\usepackage{mathtools}
\usepackage[mathcal]{euscript}
\usepackage{placeins}
\usepackage[hidelinks]{hyperref}
\usepackage{verbatim}
\usepackage[lined]{algorithm2e}
\usepackage{tikz}

\usepackage {subcaption}

\usepackage {multicol}

\usepackage {float}

\pgfplotsset{compat=newest}

\newlength{\parin}
\usepackage {array}
\newcolumntype{L}[1]{>{\raggedright\let\newline\\\arraybackslash\hspace{0pt}}m{#1}}
\newcolumntype{C}[1]{>{\centering\let\newline\\\arraybackslash\hspace{0pt}}m{#1}}
\newcolumntype{R}[1]{>{\raggedleft\let\newline\\\arraybackslash\hspace{0pt}}m{#1}}

\usetikzlibrary{decorations.pathreplacing,angles,quotes}
\usetikzlibrary{pgfplots.groupplots}
\usetikzlibrary{shapes}
\tikzstyle{fillborder} = [line width = 0.01pt, domain = 0:10, samples = 100]

\definecolor {navyblue} {rgb} {0.00, 0.00, 1.00}
\definecolor {darkred}  {rgb} {1.00, 0.00, 0.00}

\title{\textbf{Information Propagation in Stochastic Networks}}

\author{
P\'eter L. Juh\'asz \\
\scriptsize {peter@ljuhasz.com}
}

\date{\vspace{-4ex}}

\makeatletter
\renewcommand{\Indentp}[1]{%
  \advance\leftskip by #1
  \advance\skiptext by -#1
  \advance\skiprule by #1}%
\renewcommand{\Indp}{\algocf@adjustskipindent\Indentp{\algoskipindent}}
\renewcommand{\Indm}{\algocf@adjustskipindent\Indentp{-\algoskipindent}}
\makeatother

\SetKwProg{F}{}{}{}

\begin {document}

\maketitle

\FloatBarrier

\vspace{0.1cm} \begin {center} \begin {minipage} {0.7\textwidth}

  \subsubsection* {Highlights}

  \begin {itemize} \setlength \itemsep{0cm}
  \item {New temporal information propagation model for random networks}
  \item {Improved accuracy compared to the network-based SI mean-field model}
  \item {Numerical solution of the governing differential equation system}
  \item {Monte Carlo network simulation justifies the model}
  \item {Implemented simulation software is publicly available}
  \end {itemize}

\end {minipage} \end {center} \vspace{0.5cm}

\begin {multicols} {2}

\FloatBarrier
\subsubsection* {Abstract}

In this paper, a network-based stochastic information propagation model is developed.
The information flow is modeled by a probabilistic differential equation system.
The numerical solution of these equations leads to the expected number of informed nodes as a function of time and reveals the relationship between the degrees of the nodes and their reception time.
The validity of the model is justified by Monte Carlo network simulation through the analysis of information propagation in scale-free and Erd\H os--R\'enyi networks.
It has been found that the developed model provides more accurate results compared to the widely used network-based SI mean-field model, especially in sparse networks.

\FloatBarrier
\subsubsection* {Keywords}

\foreignlanguage{nohyphenation}{\begin{sloppypar}information propagation; complex system; complex network; stochastic model; SI model; Monte Carlo simulation\end{sloppypar}}

% MSC codes: https://cran.r-project.org/web/classifications/MSC.html
% 05C80: Random graphs
% 60G25: Prediction theory
% 90B15: Network models, stochastic
% 90C35: Programming involving graphs or networks
% \MSC 90B15 (Primary) \sep 05C80 \sep 60G25 \sep 90C35

\FloatBarrier
\section {Introduction}

% propagation of a piece of information
When a source node in a network creates a piece of information, it propagates through the network from node to node.
This paper aims at determining the number of informed nodes and their degree distribution as a function of time based on the stochastic modeling of the network connections.

% network definition
According to graph theory, networks are defined by an ordered triple $(V,E,\psi)$ consisting of a nonempty set $V$ of vertices, a set $E$ of edges and an incidence function $\psi$ that associates each edge with an unordered pair of vertices.\cite {bondy1976graph}
In this paper, the notions of vertex and node, and the notions of edge and connection are used interchangeably.
The degree of a node equals the number of its connections.

% Poisson process as the model of propagation
Poisson process is a widely used stochastic approach for modeling the times at which independent, random events occur.\cite {gallager2013stochastic}
It is assumed that every node in the network sends messages to each of its neighbors according to independent Poisson processes with identical parameter $\mu$.
A Poisson process is defined by a sequence of consecutive, independent, exponentially distributed interarrival times.
This means that if a neighbor of a node is informed, the time necessary for the information to reach the node is described by an exponential random variable.\footnote {
Exponential random variables possess the ``memoryless'' property, in other words, given that the waiting time for the next event is greater than a certain value, the probability distribution of the additional waiting time is the same as the unconditional distribution of the random variable.}

% external internal connections
The connections of the network are divided into two categories.
``Internal'' connections have both of their endpoints in the set of informed nodes or in the set of not informed nodes.
``External'' connections on the other hand connect an informed and a not informed node; these connections are responsible for the information propagation.

% basic idea
The next receiver node can be informed through any of the $K_{\textrm {ext}}$ external connections.
This means that the waiting time until the next node is informed is described by the minimum of independent exponential random variables with parameter $\mu$, which is also an exponential random variable (with parameter $K_{\textrm {ext}} \, \mu$).\cite {feller2008introduction}
In order to describe the dynamics of the information propagation process, the main objective is to calculate the evolution of the external connections as a function of time.

% following structure
The rest of the paper is structured as follows.
The model described in this study is compared to the related works of this field in section \ref {sec_related}.
Section \ref {sec_mathematicalmodel} discusses the assumptions on the network model and derives the differential equation system describing the information propagation.
In section {\ref {sec_connectionwiththesimodel}} the connection between the model presented in this paper and the mean-field approach of the SI model\cite {pastor2001epidemic} is detailed.
Section \ref {sec_numsol} describes the main steps of the numerical solution of the mathematical model.
The simulation algorithm is presented in section \ref {sec_simulation}.
Finally, section \ref {sec_results} presents the application of the model as well as the comparison of the simulation and the theoretical calculations.

\section {Related Work}
\label {sec_related}

Information propagation in general has been studied extensively in recent years.

% mobile communication networks
Onnela, Saram\"aki, \emph {et~al.} studied information diffusion in social communication networks through the analysis of mobile call graphs acquired from mobile phone data.\cite {onnela2007structure}
They have found that the coupling between interaction strengths and the network's local structure significantly slows the diffusion.
In the society wide communication network, the degree distribution of the nodes was skewed with a fat tail following a power law distribution (with parameter $8.4$).
This scale-free network had a giant component to which the vast majority of the nodes belong.

% email networks
Ebel \emph {et~al.} studied e-mail networks with e-mail addresses as nodes and exchanged e-mails as links.\cite {ebel2002scale}
They found that the presence of hubs accelerates the spread of computer viruses through decreasing the shortest path length between nodes.
Their research involved mainly a static analysis of the network, which had a scale free degree distribution with an exponential cutoff.
A giant component was also present in this network.

% reaction, diffusion of particles on graphs
Colizza \emph {et~al.} modeled particle diffusion in networks.\cite {colizza2007reaction}
Particles move randomly from node to node along the edges of the network.
The examined network had a power-law degree distribution which describes realistic diffusion processes well.
At each fixed length time step, every particle moves to a node in the vicinity of its current location.
Reaction of particles takes place if they are on the same node.
This model has led to a reaction--diffusion differential equation, which were verified by Monte Carlo simulations.

% book about innovations
E. M. Rogers in his famous book \emph {Diffusion of Innovations}\cite {rogers2010diffusion} studied how information flows between individuals.
Information flow tends to be more intensive through homophilious connections i.e. between similar individuals.
However, as these connections often form clusters, heterophilious connections play an important role in information flow between groups of individuals.
He also found that mass media does not have a direct effect on the opinion of individuals.
Information flow instead is better described by a two-step model, in which ``opinion leaders'' are influenced by the mass media who pass the information on to followers.

% adjacency matrix
Another approach to describe information propagation is by using matrix differential equation.\cite {mei2017dynamics}
Each row and column of the adjacency matrix $\mathbf {A}$ of the network represent a specific node.
If there is a connection between the $i$-th and $j$-th node, the $A_{ij} = 1$; $A_{ij} = 0$ otherwise.
Let us denote by $x_i (t)$ the probability that the node $i$ is informed at time $t$.
The probability that node $i$ is informed in a $\mathrm {d} t$ infinitesimal time interval, is proportional to the sum of the probabilities that its neighbors are informed.
On the other hand, $x_i$ increases only if node $i$ is not yet informed (this has a probability of $1 - x_i$).
\[ \dot {x}_i = \mu (1 - x_i) \sum \limits_j A_{ij} x_j \]
Here $\dot {x}_i$ denotes the time derivative of $x_i$.
The matrix differential equation can be written in a more compact form by introducing vector $\mathbf {x}$ containing the probabilities.
\[ \dot {\mathbf {x}} = \mu \left( \mathbf {1} - \mathrm {diag} (\mathbf {x}) \right) \mathbf {A} \mathbf {x} \]
The quadratic non-linearity poses difficulties in the solution of this equation even with several assumptions on the structure of $\mathbf {A}$.

As it can be seen from the above publications, most propagation scenarios can be described by scale-free networks with a giant component present.
The model introduced in this paper provides an asymptotically precise mathematical description of the information propagation through catching the dynamic behavior of the propagation on the connection level.
As far as the author is concerned, none of the existing network models describes the information propagation in configuration model networks more accurately than the method derived here.

\FloatBarrier
\section {Mathematical Model}
\label {sec_mathematicalmodel}

% goal, time of the next propagation step
In this section the mathematical model is developed to calculate the expected number of informed nodes as a function of time.

% construction of the network
The network is constructed according to the configuration network model, which is a widely used network model for random networks with specified degree distributions.
The first step in constructing such a network is defining the degree of each node by sampling the given degree distribution.
These degrees are represented as half links, which are paired in a random manner to form a connection.\cite {barabasi2016network}
This method enables the examination of information propagation on networks with various types of degree distributions.
Being two of the most well-known network types, special attention will be paid to Erd\H os--R\'enyi and scale-free networks.

% initial state, propagation
The propagation is initialized by choosing a source node randomly which spreads the information to its neighbors.
A state of the information propagation process is visualized in Figure \ref {fig:notations}.
As explained earlier, the information propagates from the informed nodes through $K_{\textrm {ext}}$ number of external connections connecting the informed and not informed nodes.
After the next receiver node is informed, the external connections change, which affects the speed of the propagation process.
\tikzstyle{vertex1}=[scale=1,draw,circle,fill=white!50]
\tikzstyle{edge} = [draw,-]
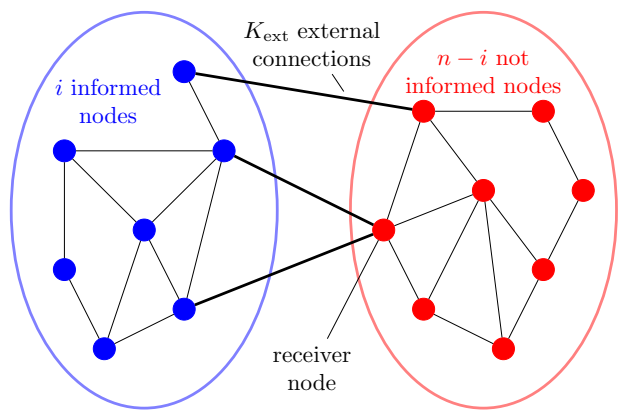
\begin {figure} [H] \centering \resizebox{0.45\textwidth}{!}{
\begin {tikzpicture} [scale=0.6, auto,swap]
  \node[ellipse, very thick, draw, minimum width=4cm, minimum height=6cm, opacity=0.5, navyblue] at (2, 3.5) {};
  \node[ellipse, very thick, draw, minimum width=4cm, minimum height=6cm, opacity=0.5, red] at (10.5, 3.5) {};

  \node[vertex1, navyblue] at  (0,2)  (1) {};
  \node[vertex1, navyblue] at  (0,5)  (2) {};
  \node[vertex1, navyblue] at  (1,0)  (3) {};
  \node[vertex1, navyblue] at  (2,3)  (4) {};
  \node[vertex1, navyblue] at  (3,1)  (5) {};
  \node[vertex1, navyblue] at  (4,5)  (6) {};
  \node[vertex1, navyblue] at  (3,7)  (7) {};

  \node[vertex1, red] at  (8,3)    (8) {};
  \node[vertex1, red] at  (9,1)    (9) {};
  \node[vertex1, red] at  (9,6)   (10) {};
  \node[vertex1, red] at (10.5,4) (11) {};
  \node[vertex1, red] at (11,0)   (12) {};
  \node[vertex1, red] at (12,2)   (13) {};
  \node[vertex1, red] at (12,6)   (14) {};
  \node[vertex1, red] at (13,4)   (15) {};

  \path[edge]  (1) --  (2);
  \path[edge]  (1) --  (3);
  \path[edge]  (3) --  (4);
  \path[edge]  (3) --  (5);
  \path[edge]  (4) --  (5);
  \path[edge]  (4) --  (6);
  \path[edge]  (2) --  (4);
  \path[edge]  (2) --  (6);
  \path[edge]  (5) --  (6);
  \path[edge]  (6) --  (7);

  \path[edge]  (8) --  (9);
  \path[edge]  (8) -- (10);
  \path[edge]  (8) -- (11);
  \path[edge]  (9) -- (11);
  \path[edge]  (9) -- (12);
  \path[edge] (10) -- (11);
  \path[edge] (10) -- (14);
  \path[edge] (11) -- (12);
  \path[edge] (11) -- (13);
  \path[edge] (12) -- (13);
  \path[edge] (13) -- (15);
  \path[edge] (14) -- (15);

  \path[very thick, edge]  (5) --  (8);
  \path[very thick, edge]  (6) --  (8);
  \path[very thick, edge]  (7) -- (10);

  \node [navyblue, align=center] at (1.1, 6.25) {$i$ informed\\nodes};
  \node [red, align=center] at (10.5, 7) {$n - i$ not\\informed nodes};

  \node [align=center] at (6.2, 7.7) (kext) {$K_{\textrm {ext}}$ external\\connections};
  \path [edge] (kext) -- (7.0, 6.5);
  % \path [edge] (kext) -- (5.5, 4.5);
  % \path [edge] (kext) -- (3.5, 1.3);

  \node [align=center] at (6.2, -0.5) (recv) {receiver\\node};
  \path [edge] (recv) -- (8);

\end {tikzpicture}}
\caption {
  At a given point in time there are $i$ informed nodes (left) and $n-i$ not informed nodes (right).
  Information can propagate through the $K_{\textrm {ext}}$ external connections that connect informed and not informed nodes.
  The receiver node has a degree of $k_{\textrm {recv}}$; out of these $k_{\textrm {recv}}^{\textrm {inf}}$ connect the receiver node with the informed nodes.}
  \label {fig:notations}
\end {figure}

% notations of variables
\begin {table} [H] \centering \begin {tabular} {lL{6.5cm}}
$n$                                  & total number of nodes in the network \\
$i$                                  & number of informed nodes \\
$t$                                  & elapsed time since the first node was informed \\
$\mu$                                & characteristic time of the information propagation \\
$k_{\textrm {tot}}$                  & degree of a randomly chosen node \\
$k_{\textrm {inf}}$                  & degree of a randomly chosen informed node \\
$k_{\textrm {ninf}}$                 & degree of a randomly chosen not informed node \\
$k_{\textrm {max}}$                  & maximal degree in the network \\
$k_{\textrm {recv}}$                 & degree of the receiver node \\
$k_{\textrm {recv}}^{\textrm {inf}}$ & number of connections between the informed nodes and the receiver node \\
$K_{\textrm {tot ni}}$               & sum of degrees belonging to the not informed nodes \\
$K_{\textrm {ext}}$                  & number of external connections between the informed and not informed nodes
\end {tabular} \caption {Notations used in the mathematical derivation} \label {tab:notations} \end {table}
Next, let us introduce some notations.
Table \ref {tab:notations} contains the definitions of the quantities used throughout the paper.
Apart from $n$, $t$, $\mu$ and $k_{\textrm {tot}}$, these quantities are time-dependent random variables (i.e. stochastic processes).
The time dependence is not denoted explicitly for the sake of clarity.
While quantities denoted by normal ``$k_{\cdots}$'' fonts refer to the connections of one node, notations with capital ``$K_{\cdots}$'' fonts refer to aggregate quantities over a set of nodes.

% notations for distributions
In general, the probability mass functions of the random variables depend on the number of informed nodes.
The probability mass function of a random variable $X$ at the time when there are $i$ informed nodes is denoted by $P_X (x \, | \, i)$ with the below definition.
\[ P_X (x \, | \, i) = \mathbb {P} \left( X = x \, | \, \textrm {number of informed nodes} = i \right) \]
The expectation of any random variable is denoted by $\mathbb {E} [\cdot]$ throughout this paper.
$\mathbb {E}^n [\cdot]$ is the shorthand form of $(\mathbb {E} [\cdot])^n$.
The variance of a random variable is denoted by $\mathbb {D}^2 [\cdot]$.

\subsubsection* {Differential Equation of the Propagation}

% goal, time of the next propagation step
We would like to calculate the distribution of the number of informed nodes $i$ as a function of time.

Suppose that currently $i$ number of nodes are informed.
As it was discussed earlier, the distribution of the time interval $\Delta t$ until the next (uninformed) node is informed is the minimum of $K_{\textrm {ext}}$ number of independent exponential random variables with parameter $\mu$.
The minimum of independent exponential random variables is again an exponential random variable with parameter $\mu K_{\textrm {ext}}$.
If we define $\Delta t$ as the time until the next $\Delta i$ number of nodes is informed, the below distribution holds.
\[ \frac {\Delta t} {\Delta i} \sim \mathrm {Exponential} \left( \mu K_{\textrm {ext}} \right) \]
Note that $\Delta t / \Delta i$ is a random variable, and the term ``$\mathrm {Exponential}$'' refers to its distribution.
The expectation of $\Delta t / \Delta i$ is the following.
\begin {equation} \mathbb {E} \left[ \frac {\Delta t} {\Delta i} \right] = \frac {1} {\mu} \mathbb {E} \left[ \frac {1} {K_{\textrm {ext}}} \right] \label {eq:base} \end {equation}
Using summation, the expectation of the time $t(i)$ can be given as follows.
\[ \mathbb {E} t = \mathbb {E} \left[ \sum \Delta t \right] = \sum \mathbb {E} \left[ \Delta t \right] = \sum \mathbb {E} \left[ \frac {\Delta t} {\Delta i} \right] \, \Delta i \]

% propagation can stop, interconnected network, giant component
Note that the expectation of $\Delta t / \Delta i$ is not bounded as there is a finite probability that $K_{\textrm {ext}}$ is zero.
In this case there are no external connections any more so new nodes cannot be informed and the propagation is interrupted (stopped).
In other words there are two separate issues we have to deal with.
\begin {itemize}
\item {What is the probability that $i$ number of nodes will be informed?
\[ \prod_{j = 1}^i \left( 1 - P_{K_{\textrm {ext}}} (0 \, | \, j) \right) \]
This is the probability that the component in which the propagation is started contains at least i nodes.}
\item {Provided that $i$ number of nodes will be informed, what is the expectation of the time $t(i)$, when the $i$-th node is informed?}
\end {itemize}
If the propagation starts in a small component, $P_{K_{\textrm {ext}}} (0 \, | \, i)$ is high at the beginning.
On the other hand, $P_{K_{\textrm {ext}}} (0 \, | \, i)$ is also high towards the end of the propagation, when most of the nodes are informed in the component.
However, throughout the majority of the propagation process this probability is negligible in not too sparse networks.
In this case the network contains a connected giant component, in which all of the nodes are reachable from every other node.
Let us denote by $K_\textrm {tot}$ the degree of a randomly chosen node from the whole network.
According to the Molloy--Reed criterion\cite{molloy1995critical}, the condition of the existence of a giant component is the following.
\[ \mathbb {E} K_\textrm {tot}^2 - 2 \mathbb {E} K_\textrm {tot} > 0 \]
When choosing the parameters for the degree distributions, we always pay attention to the above criterion.\footnote {Note that the network cannot be interconnected if the average degree of the network is smaller than $2$, i.e., if $\mathbb {E} k_{\textrm {tot}} < 2$.}

% method of moments
In order to approximate the expectation in \eqref {eq:base}, the method of moments is used, which is based on the Taylor-series expansion around the expected value.\cite {matyas1999generalized}
The precondition of this approach is that the probability of $K_{\textrm {ext}} = 0$ is neglected.
\[ \mathbb {E} \left[ \frac {1} {K_{\textrm {ext}}} \right] \approx \frac {1} {\mathbb {E} K_{\textrm {ext}}} \left( 1 + \frac {\mathbb {D}^2 K_{\textrm {ext}}} {\mathbb {E}^2 K_{\textrm {ext}}} + \cdots \right) \]
$K_{\textrm {ext}}$ is the sum of multiple uncorrelated random variables, therefore $\mathbb {D}^2 K_{\textrm {ext}} / \mathbb {E}^3 K_{\textrm {ext}}$ is negligible.
In this case the $\mathbb {E} [ 1 / K_{\textrm {ext}} ] \approx 1 / \mathbb {E} K_{\textrm {ext}}$ approximation can be used.

% continuous limit
Taking the continuous limit of \eqref {eq:base} can simplify the handling of the differential equation of the expectation by widening the range of the applicable numerical methods without a significant decrease in accuracy.
\begin{equation} \frac {\mathrm {d} \, \mathbb {E} t} {\mathrm {d} i} = \frac {1} {\mu} \frac {1} {\mathbb {E} K_{\textrm {ext}}} \label {eq:dtdi} \end{equation}

\subsubsection* {Evolution of the External Connections}

% goal: external connections
In order to solve \eqref {eq:dtdi}, the expectation $\mathbb {E} K_{\textrm {ext}}$ has to be determined as a function of the number of informed nodes~$i$.
$K_{\textrm {ext}}$ (and its expectation) changes whenever a new node -- the receiver -- is informed.

% receiver node and new connections (delta K_ext)
Out of the $k_{\textrm {recv}}$ connections of the next receiver node, $k_{\textrm {recv}}^{\textrm {inf}}$ connects the receiver node with the previously informed nodes.
As for each of these another half link belongs to the previously informed nodes, $2 k_{\textrm {recv}}^{\textrm {inf}}$ connections become internal in total.
This means that the change of the external connections $\Delta K_{\textrm {ext}}$ is the following.
\[ \Delta K_{\textrm {ext}} = k_{\textrm {recv}} - 2 k_{\textrm {recv}}^{\textrm {inf}} \]
These calculations can be extended for $\Delta i$ new informed nodes.
\[ \frac {\Delta K_{\textrm {ext}}} {\Delta i} = k_{\textrm {recv}} - 2 k_{\textrm {recv}}^{\textrm {inf}} \]
Taking the expectation of $\Delta K_{\textrm {ext}}$, and the continuous limit, the following formula results.
\begin {equation} \frac {\mathbb {E} \left[ \Delta K_{\textrm {ext}} \right]} {\Delta i} = \mathbb {E} k_{\textrm {recv}} - 2 \mathbb {E} k_{\textrm {recv}}^{\textrm {inf}}
\label {eq:k_ext} \end {equation}

% conditional expectation of internal connections
Let us calculate $\mathbb {E} k_{\textrm {recv}}^{\textrm {inf}}$, the expected number of connections between the receiver node and the informed nodes.
Altogether there are $K_{\textrm {ext}} + K_{\textrm {tot ni}}$ half links to which each of the not informed nodes can connect to, so the distribution of $k_{\textrm {recv}}^{\textrm {inf}}$ is related to the binomial distribution with parameters $k_{\textrm {recv}}$ and $p = K_{\textrm {ext}} / (K_{\textrm {ext}} + K_{\textrm {tot ni}})$.\footnote {
As the receiver node cannot be connected to a single half link with multiple connections, the distribution of $k_{\textrm {recv}}^{\textrm {inf}}$ follows a hypergeometric distribution.
However, for large networks (when $K_{\textrm {ext}} + K_{\textrm {tot ni}} \gg k_{\textrm {recv}}$), the hypergeometric distribution is well approximated by the binomial distribution.}
It also has to be taken into account that the more connections a not informed node has with the informed nodes, the higher the probability that it will be the next receiver node.
In other words -- provided that $k_{\textrm {recv}}$, $K_{\textrm {ext}}$ and $K_{\textrm {tot ni}}$ are fixed --, $k_{\textrm {recv}}^{\textrm {inf}}$ follows a weighted binomial distribution with the below probability mass function, conditioned on the values of $k_{\textrm {recv}}, K_{\textrm {tot ni}}, K_{\textrm {ext}}$.
\[ \begin {aligned}
P_{k_{\textrm {recv}}^{\textrm {inf}} \, | \, k_{\textrm {recv}}, K_{\textrm {tot ni}}, K_{\textrm {ext}}} \left( k \right) &\sim k \, {k_{\textrm {recv}} \choose k} \, p^k \, \left( 1 - p \right)^{k_{\textrm {recv}} - k} \\
p &= \frac {K_{\textrm {ext}}} {K_{\textrm {ext}} + K_{\textrm {tot ni}}}
\end {aligned} \]
Note that the normalization constant was omitted for clarity.
It follows that the conditional expectation of $k_{\textrm {recv}}^{\textrm {inf}}$ is given by the below formula.
\[ \begin {aligned}
\mathbb {E} &\left[ k_{\textrm {recv}}^{\textrm {inf}} \, | \, k_{\textrm {recv}}, \, K_{\textrm {ext}}, \, K_{\textrm {tot ni}} \right] = \\
&= 1 + \left( k_{\textrm {recv}} - 1 \right) \, \frac {K_{\textrm {ext}}} {K_{\textrm {ext}} + K_{\textrm {tot ni}}}
\end {aligned} \]

% unconditional expectation
To calculate the total (i.e. unconditional) expectation of $k_{\textrm {recv}}^{\textrm {inf}}$, Taylor--series expansion around the expected values is used again,\cite {matyas1999generalized} and only the first order term is kept.
\begin {equation} \mathbb {E} k_{\textrm {recv}}^{\textrm {inf}} \approx 1 + \left( \mathbb {E}k_{\textrm {recv}} - 1 \right) \, \frac {\mathbb {E} K_{\textrm {ext}}} {\mathbb {E} K_{\textrm {ext}} + \mathbb {E} K_{\textrm {tot ni}}} \label {eq:krecvint} \end {equation}

% approximation error
Keeping only the first order term means neglecting the covariance between the variables.
The neglected terms being positive, the number of external connections is overestimated in \eqref {eq:k_ext}.
The integration of \eqref {eq:k_ext} thus leads to increasing approximation error for later stages of the propagation.

% spanning tree
This result clearly shows that at least one connection connects the receiver node to the informed nodes, which ensures that the information could reach the receiver node.
It follows that a spanning tree connects all the informed nodes, which corresponds to the intuition.
The rest of the $k_{\textrm {recv}} - 1$ connections are distributed randomly among the free half links.

\subsubsection* {Degree Distribution of the Informed Nodes}

The final task is to determine the degree distributions of the receiver and not informed nodes in order to calculate the expected values $\mathbb {E} k_{\textrm {recv}}$ and $\mathbb {E} K_{\textrm {tot ni}}$.

% degree distributions
The $n$ number of nodes in the network is assumed to be a known constant.
As mentioned earlier, the degrees of a randomly chosen informed and not informed node are denoted by $k_{\textrm {inf}}$ and $k_{\textrm {ninf}}$ respectively.
%The degree distributions of the informed and not informed nodes are denoted by $P_{k_{\textrm {inf}}} (k \, | \, i)$ and $P_{k_{\textrm {ninf}}} (k \, | \, i)$ respectively, where the distributions depend on the number of informed nodes $i$.

% degree distribution of the receiver
The probability that a randomly chosen node is connected to a node is proportional to its degree.
This means that the degree distribution of the receiver node $P_{k_{\textrm {recv}}}$ is the weighted degree distribution of the not informed nodes $P_{k_{\textrm {ninf}}}$.
\begin {equation} P_{k_{\textrm {recv}}} (k \, | \, i) = \frac {k} {\mathbb {E} k_{\textrm {ninf}}} P_{k_{\textrm {ninf}}} (k \, | \, i) \label {eq:recv} \end {equation}

% new degree distribution of the not informed nodes
The ``mixture'' of the new degree distribution $P_{k_{\textrm {ninf}}} (k \, | \, i + 1)$ (represented by $n - i - 1$ nodes) and the degree distribution of the receiver node $P_{k_{\textrm {recv}}} (k \, | \, i)$ is the original degree distribution $P_{k_{\textrm {ninf}}} (k \, | \, i)$.
\[ \begin {aligned}
P_{k_{\textrm {ninf}}} (k \, | \, i) &= \frac {\left( n - i - 1 \right) \, P_{k_{\textrm {ninf}}} (k \, | \, i + 1)} {n - i} + \\
&+ \frac {P_{k_{\textrm {recv}}} (k \, | \, i)} {n - i} \\
P_{k_{\textrm {ninf}}} (k \, | \, i + 1) &= \frac {\left( n - i \right) \, P_{k_{\textrm {ninf}}} (k \, | \, i) - P_{k_{\textrm {recv}}} (k \, | \, i)} {n - i - 1} \\
\frac {\partial P_{k_{\textrm {ninf}}} (k \, | \, i)} {\partial i} &= \frac {P_{k_{\textrm {ninf}}} (k \, | \, i) - P_{k_{\textrm {recv}}} (k \, | \, i)} {n - i}
\end {aligned} \]
The continuum formalism is introduced as the limit of $\Delta i \to 0$ is considered.
This limit is valid, if $i \gg 1$, i.e., when the calculations consider several informed nodes.\footnote {
As the differential equations are solved using numerical methods, the magnitude of the $\Delta i$ steps can easily be ensured to be at least $1$, which means that the $\Delta i \to 0$ limit is not considered in practice.}
If the degree distribution of the receiver node is substituted into this formula, the below differential equation system follows.
\begin {equation} \frac {\partial P_{k_{\textrm {ninf}}} (k \, | \, i)} {\partial i} = \frac {P_{k_{\textrm {ninf}}} (k \, | \, i)} {n - i} \left( 1 - \frac {k} {\mathbb {E} k_{\textrm {ninf}}} \right) \label {eq:ninfdegdist} \end {equation}
This is a differential equation system for all values of $k$.
Note that the total degree distribution of the nodes as well as the initial degree distribution is arbitrary.
Solving this system of equations we can determine the degree distribution of the not informed nodes as a function of the number of informed nodes.

% degree distribution of the informed nodes
The degree distribution of the not informed nodes can be calculated in a similar way.
The total degree distribution $P_{k_{\textrm {tot}}} (k)$ is the mixture of the degree distributions of the informed nodes $P_{k_{\textrm {inf}}} (k \, | \, i)$ (represented by $i$ nodes) and not informed nodes $P_{k_{\textrm {ninf}}} (k \, | \, i)$ (represented by $n - i$ nodes).
\begin {equation} \begin {aligned}
  P_{k_{\textrm {tot}}} (k) &= \frac {1} {n} \left[ i \, P_{k_{\textrm {inf}}} (k \, | \, i) + (n - i) P_{k_{\textrm {ninf}}} (k \, | \, i) \right] \\
  P_{k_{\textrm {inf}}} (k \, | \, i) &= \frac {1} {i} \left[ n \, P_{k_{\textrm {tot}}} (k) - (n - i) P_{k_{\textrm {ninf}}} (k \, | \, i) \right]
\label {eq:infdegdist}
\end {aligned} \end {equation}

\subsubsection* {Equations of the Propagation}

To sum up, the steps of the solution are the following.
\begin {itemize}
\item {First, \eqref {eq:ninfdegdist} has to be solved to get the degree distribution of the not informed nodes.
The solution of \eqref {eq:recv} results in the degree distribution of the next receiver node.}
\item {Next, \eqref {eq:k_ext} leads to the expected number of external connections.}
\item {Finally, by integrating \eqref {eq:dtdi} and taking its inverse, the expected number of informed nodes as a function of time is given.}
\end {itemize}

\section {Connection With the Mean-Field Model}
\label {sec_connectionwiththesimodel}

% introduction: mean-field model
R. Pastor-Satorras and A. Vespignani described epidemic spreading in scale-free networks.\cite {pastor2001epidemic}
They analyzed the spread of computer viruses on the Internet by defining a model, which is based upon dynamical mean-field reaction rate equations.
The epidemic spreading was modeled by the susceptible-infected-susceptible (SIS) model, according to which each node is in either susceptible or infected state.
Next it is shown how the degree-based mean-field approach of the SI model is related to the stochastic model presented above.

% degree-based mean-field equation
A node can become infected with rate $\nu$ if it is connected to at least one infected node.
Infected nodes become susceptible again with rate $\delta$.
The effective spreading rate is the ratio $\mu_{\textrm {SIS}} := \nu / \delta$.
Let $\rho_k (t)$ denote the time dependent probability that a randomly chosen node with a degree of $k$ is infected.
The degree distribution of the nodes is $P_{k_{\textrm {tot}}} (k)$.\footnote {The notations used in the original paper were changed so that they are consistent with the ones used here.}
In this case the differential equation driving the dynamics of the epidemic is as below.
\[ \partial_t \rho_k (t) = - \rho_k (t) + \mu_\textrm {SIS} k \left( 1 - \rho_k (t) \right) \, \frac {\sum_{k'} \rho_{k'} (t) k' \lambda_\textrm {tot} (k')} {\sum_{k''} k'' \lambda_\textrm {tot} (k'')} \]
If the state transition from infected to susceptible is not possible, the equations valid for the resulting susceptible-infected (SI) model does not contain the first negative term, thus $\mu_\textrm {SI} = \nu$.
\[ \partial_t \rho_k (t) = \mu_\textrm {SI} k \left( 1 - \rho_k (t) \right) \, \frac {\sum_{k'} \rho_{k'} (t) k' P_{k_{\textrm {tot}}} (k')} {\sum_{k'} k' P_{k_{\textrm {tot}}} (k')} \]
At any given time $t$, the number of infected nodes with degree $k$ is $i(t) \, P_{k_{\textrm {inf}}} (k \, | \, i(t))$.
The total number of nodes with degree $k$ is $n \, P_{k_{\textrm {tot}}} (k)$, which means that the probability that a node with degree $k$ is infected is given by the below formula.
\[ \rho_k (t) = \frac {i(t) \, P_{k_{\textrm {inf}}} (k \, | \, i(t))} {n \, P_{k_{\textrm {tot}}} (k)} \]
Let us transform the partial differential equation for the degree distributions \eqref {eq:infdegdist} first.
\[ \begin {aligned}
\frac {\partial \left( i \, P_{k_{\textrm {inf}}} (k \, | \, i) \right)} {\partial i} &= \frac {n P_{k_{\textrm {tot}}} (k) - i P_{k_{\textrm {inf}}} (k \, | \, i)} {S - i \mathbb {E} k_{\textrm {inf}}} k \\
\frac {\partial \rho_k (i)} {\partial i} &= \frac {1 - \rho_k (i)} {S - i \mathbb {E} k_{\textrm {inf}}} k
\end {aligned} \]
This means, that \eqref {eq:infdegdist} can be transformed into the respective equation for the mean-field model.

% SI model as a baseline
The rest of the propagation is handled in a different manner.
For instance, the internal connections used for the propagation and their weighted distribution are not taken into account.
These relationships motivate to evaluate the accuracy of our model using the mean-field model as a baseline.

\FloatBarrier
\section {Numerical Solution}
\label {sec_numsol}

% construction of Erdos-Renyi and scale-free networks
Let us examine the construction of Erd\H os--R\'enyi and scale-free networks.
A network is said to be scale-free if its degree distribution follows a power law.\cite {onnela2007structure}
An Erd\H os--R\'enyi network is constructed by first fixing the $n$ number of nodes in the network, an then connecting each pair of nodes with a given probability $p$.\cite {gilbert1959random}
This leads to a binomial degree distribution.
If the expected number of node degree $n p$ is fixed as the number of nodes increases, the degree distribution is described by a Poisson distribution for large $n$ values.
The minimum degree value $k_{\textrm {min}}$ must be chosen in such a way so that a giant component exists.

% maximum of k
Regarding the maximum degree value $k_{\textrm {max}}$ it can be expected that the probability that a node has a degree higher than $k_{\textrm {max}}$ is less than $1/n$.\cite {barabasi2016network}
This $k_{\textrm {max}}$ value is called the ``natural cutoff'' of the degree distribution.
\[ \int \limits_{k_{\textrm {max}}}^\infty P_{k_{\textrm {tot}}} (k) \, \mathrm {d} k = \frac {1} {n} \]
For Erd\H os--R\'enyi and scale free networks, $k_{\textrm {max}}$ is given as follows.
\[ \begin {aligned}
k_{\textrm {max}}^{\textrm {ER}} &= \mathrm {e}^\gamma \, \left( 1 - \frac {1} {n} \right) \\
k_{\textrm {max}}^{\textrm {SF}} &= k_{\textrm {min}} \, n^{\frac {1} {\gamma' - 1}}
\end {aligned} \]
Here $\gamma$ and $\gamma'$ refer to the parameters of the degree distributions.
Note that in the case of scale-free networks, the maximum degree does not exist whenever the parameter $\gamma' \leq 2$.

% expected value
By integration, the expectations of the node degrees for Erd\H os--R\'enyi and scale-free networks are given as follows.
\[ \begin {aligned}
\mathbb {E} k_{\textrm {tot}}^{\textrm {ER}} &= \gamma \\
\mathbb {E} k_{\textrm {tot}}^{\textrm {SF}} &= \frac {\gamma' - 1} {\gamma' - 2} \, \frac {k_{\textrm {min}}^{2 - \gamma'}} {k_{\textrm {min}}^{1 - \gamma'}} \, \frac {1 - n^{- \frac {\gamma' - 2} {\gamma' - 1}}} {1 - n^{-1}}
\end {aligned} \]

% initial conditions
The propagation starts from a randomly chosen informed node.
The time measurement is started after a defined fraction of the nodes is informed so that the approximation in \eqref {eq:dtdi} holds.

% Newton method
The differential equations derived in section \ref {sec_mathematicalmodel} are integrated numerically.
Numerical integration was implemented using the low-resource Newton method, enabling the solution in a reasonable time-frame.

\begin {figure} [H] \centering
  \begin{tikzpicture}
    \begin{axis}[
        axis x line=middle,
        xmode = log,
        axis y line=none,
        height=3cm,
        width=\axisdefaultwidth,
        xmin=0.5,
        xmax=1000,
        xtick={1, 10, 100, 500},
        xticklabels={$i_0$, $10 i_0$, $100 i_0$, $n$},
        minor tick style={draw=none},
        scaled x ticks=false,
        xlabel = $\mathrm {log}(i)$,
        every axis x label/.style={at={(current axis.right of origin)},anchor=west},
      ]
      \addplot [mark=none] coordinates{(1,0)};
      \draw    [decoration={brace, raise=5pt},decorate] (1  ,0) -- node[above=6pt] {$\#_{\Delta i}$ steps} (10  ,0);
      \draw    [decoration={brace, raise=5pt},decorate] (10 ,0) -- node[above=6pt] {$\#_{\Delta i}$ steps} (100 ,0);
      \addplot [white, mark=none] coordinates{(104,0) (300,0)};
      \addplot [mark=*, only marks, mark size=0.75] coordinates{(131.61,0) (173.21,0) (227.95,0)};
    \end{axis}
  \end{tikzpicture}
  \caption {
    The strategy of determining the length of $\Delta i$ for the numerical integration.
    The distinct sections contain $\#_{\Delta i}$ equidistant steps.
    } \label {fig:intervals}
\end {figure}
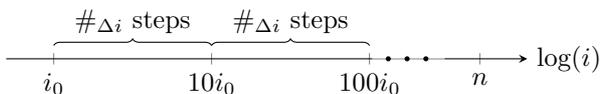

% Delta i is increasing for larger i values
Although the numerical solution of these equations is simple for small networks, it is not feasible to calculate the degree distributions $P_{K_{\textrm {tot inf}}} (k \, | \, i)$ of the informed nodes for every value of $i$ for larger networks.
As the solution is more sensitive to this value when the number of informed nodes $i$ is small, the following strategy is used to define the size of the steps.
The interval between the initial number of informed nodes $i_0$ and the total number of nodes $n$ is divided into sections.
The upper limit of the subsequent sections is increasing by a magnitude of $1$, i.e., their length is increasing $9$ times compared to the previous section.
The number of steps used in each section is denoted by $\#_{\Delta i}$.
This strategy is visualized in Figure \ref {fig:intervals}.

\FloatBarrier
\section {Simulation}
\label {sec_simulation}

% concept: ensemble of networks, then statistics
The results of the mathematical model are verified by event-based Monte Carlo simulations through handling an ensemble of random networks.
The logging of the reception times enables the calculation of various statistics.

% event driven simulation
The core element of the event-based simulation is the event queue, which contains a list of events ordered by their time.
Each event represents one step of the propagation consisting of a sending time and a target node.

% programming language, multi threading
C++ was chosen as the programming language for the implementation.
Being a computation heavy process, the simulation utilizes multiple threads to increase efficiency.
The implementation is available at GitHub.\cite {simulator}

% build network
The \emph {build network} procedure constructs the network according to the configuration model.
First degrees are assigned to the nodes based on the degree distribution.
After the degrees of the nodes are created, randomly chosen chunks are connected as long as it is possible to create additional connections.

% initial conditions
The \emph {initialize propagation} function chooses a random node and a new event is created for each of its neighbors.
These events are then put into the event queue, forming the first events of the propagation.

% simulation
During the \emph {simulate propagation} method the events are processed, and a global time variable tracks the time according to the time attribute of the subsequent event.
If the receiver node of the next event is already informed, the simulation continues.
In each step, new events are added to the event queue for each neighbor of the newly informed node.
The simulation continues until the event queue becomes empty.

% main
%   load configuration
%   initialize logger
%   
%   job
%     while (available simulations to run)
%       network = build network
%         create builder
%         builder.construct
%           nodes = create nodes
%           create degree distribution
%           assign degrees to nodes
%           connect nodes randomly
%             create list of free degrees
%             while (possible to create new connections)
%               create random connection
%       event queue = initialize propagation
%         create event queue
%         inject information to a random node
%         create events for each neighbors in event queue
%       run_simulation
%         while (event queue is not empty)
%           get next event
%           if (next event target node is not informed)
%             inform target node
%             get list of neighbors
%             update event queue with new events
%       save result of current simulation
%   export results

% algorithm pseudo code
The main procedures of the algorithm are illustrated by Algorithm \ref {sim:main} pseudocode.
\begin {figure} [H] \centering
\begin {algorithm}[H]
  \While {remaining simulations} {
    \F {\textbf {build network}} {
      create nodes\;
      assign degrees to nodes\;
      
      \F {\textbf {connect nodes randomly}} {
        create array of free degrees\;
        \While {new connections possible} {
          create random connection\;
        }
      }
    }
    \F {\textbf {initialize propagation}} {
      create event queue\;
      inject information to a random node\;
      create new events for each neighbor\;
    }
    \F {\textbf {simulate propagation}} {
      \While {event queue is not empty} {
        get next event\;
        adjust time based on next event\;
        \If {target node is not informed} {
          inform target node\;
          get list of neighbors\;
          update event queue with new events\;
        }
      }
    }
    \F {\textbf {save results}} {
    }
  }
\caption {Pseudocode of Simulation} \label {sim:main}
\end {algorithm}
\end {figure}

\FloatBarrier
\section {Results}
\label {sec_results}

% analysis of scale-free and Erdos-Renyi networks
In this section, a detailed analysis is provided for the accuracy of our model in scale-free and Erd\H os--R\'enyi networks.

% network parameters
The number of nodes in the networks is $10\,000$, and in each case $10\,000$ independent simulations were executed to approximate the results.
As the approximation made in \eqref {eq:dtdi} is inaccurate when the number of informed nodes is small, the initial instant of the analysis is the time when $5$ nodes are informed.
The parameter $\mu$ is set to be $1$ for all of the simulations and calculations, as it only affects the timescale of the propagation.
In order to fulfill the Molloy--Reed criterion, the minimum degrees of the power law and Poisson degree distributions is $2$ and $1$ respectively.

% time measurement, time unit
Equation \eqref {eq:dtdi} shows that the parameter $\mu$ controls the time scale of the propagation through determining the frequency of the ``sending events'' on each connection.
Being only a multiplicative factor, we can set $\mu := 1$ without the loss of generality.
The time unit of the propagation is also arbitrary for similar reasons, thus time will be measured in ``time units'' (referring to a unit in which the parameter $\mu$ has the value of $1$).

% propagation threshold
Building the network based on the configuration model does not ensure the presence of a giant component.
Furthermore, even if a giant component exists, the first informed node is not always chosen to be in that component due to the random initialization.
In order to ensure that the information propagates through the network, a threshold is taken into account: only those simulations are considered, where the propagation reaches $99\%$ of the nodes.

\subsubsection* {Performance Evaluation}

\begin {figure} [H] \centering
  \begin {subfigure} {0.45\textwidth} \centering
    \resizebox {\textwidth} {!} {
      \begin {tikzpicture}
        \begin {axis} [
          xmin=0, xmax=5.25, ymin=0, ymax=1.05,
          axis lines = middle,
          grid = both,
		   minor tick num=1,
          xlabel={time unit}, ylabel={ratio of informed nodes},
          xtick={0.005, 1, 2, 3, 4, 5},
          xticklabels={$0$, $1$, $2$, $3$, $4$, $5$},
          ytick={0.001, 0.2, 0.4, 0.6, 0.8, 1.0},
          yticklabels={$0\%$, $20\%$, $40\%$, $60\%$, $80\%$, $100\%$},
          xlabel style={anchor=north, at={(axis description cs:  0.50, -0.1)}},
          ylabel style={anchor=south, at={(axis description cs: -0.15,  0.5)}, rotate=90},
          legend pos=south east,
          legend cell align={left}
        ]
          \node [fill = white, align = left, rotate = 60] at (axis cs: 2.00, 0.6){$k^{-3.00}$};
          \node [fill = white, align = left, rotate = 68] at (axis cs: 1.30, 0.6){$k^{-2.75}$};
          \node [fill = white, align = left, rotate = 78] at (axis cs: 0.65, 0.6){$k^{-2.50}$};

          \addplot [ultra thick,         color = darkred                 ] table [x index = 0, y index = 3                               ] {propagation_pow300.dat};
          \addplot [ultra thick, dashed, color = navyblue                ] table [x index = 1, y index = 3, restrict y to domain=-1  :0.8] {propagation_pow300.dat};
          \addplot [ultra thick, dashed, color = navyblue!20, forget plot] table [x index = 1, y index = 3, restrict y to domain= 0.8:1.0] {propagation_pow300.dat};

          \addplot [ultra thick,         color = darkred                 ] table [x index = 0, y index = 4                               ] {propagation_pow275.dat};
          \addplot [ultra thick, dashed, color = navyblue                ] table [x index = 1, y index = 4, restrict y to domain=-1  :0.8] {propagation_pow275.dat};
          \addplot [ultra thick, dashed, color = navyblue!20, forget plot] table [x index = 1, y index = 4, restrict y to domain= 0.8:1.0] {propagation_pow275.dat};

          \addplot [ultra thick,         color = darkred                 ] table [x index = 0, y index = 2                               ] {propagation_pow250.dat};
          \addplot [ultra thick, dashed, color = navyblue                ] table [x index = 1, y index = 2, restrict y to domain=-1  :0.8] {propagation_pow250.dat};
          \addplot [ultra thick, dashed, color = navyblue!20, forget plot] table [x index = 1, y index = 2, restrict y to domain= 0.8:1.0] {propagation_pow250.dat};

          \addlegendentry {simulation}
          \addlegendentry {stochastic approx.}
        \end {axis}
      \end {tikzpicture}}
    \caption {Scale-free networks}
    \label {fig:evaluation_pow}
  \end {subfigure}
  \\[.5cm]
  \begin {subfigure} {0.45\textwidth} \centering
    \resizebox {\textwidth} {!} {
      \begin {tikzpicture}
          \begin {axis} [
      xmin=0, xmax=5.25, ymin=0, ymax=1.05,
      axis lines = middle,
      grid = both,
      minor tick num=1,
      xlabel={time unit}, ylabel={ratio of informed nodes},
      xtick={0.005, 1, 2, 3, 4, 5},
      xticklabels={$0$, $1$, $2$, $3$, $4$, $5$},
      ytick={0.001, 0.2, 0.4, 0.6, 0.8, 1.0},
      yticklabels={$0\%$, $20\%$, $40\%$, $60\%$, $80\%$, $100\%$},
      xlabel style={anchor=north, at={(axis description cs:  0.50, -0.1)}},
      ylabel style={anchor=south, at={(axis description cs: -0.15,  0.5)}, rotate=90},
      legend pos=south east,
      legend cell align={left}
    ]
      \node [fill = white, align = left, rotate = 60] at (axis cs: 2.70, 0.6){$\mathrm {Poisson} (3.0)$};
      \node [fill = white, align = left, rotate = 71] at (axis cs: 1.75, 0.6){$\mathrm {Poisson} (4.0)$};
      \node [fill = white, align = left, rotate = 78] at (axis cs: 1.20, 0.6){$\mathrm {Poisson} (5.0)$};

      \addplot [ultra thick,         color = darkred                 ] table [x index = 0, y index = 2                               ] {propagation_poi300.dat};
      \addplot [ultra thick, dashed, color = navyblue                ] table [x index = 1, y index = 2, restrict y to domain=-1  :0.8] {propagation_poi300.dat};
      \addplot [ultra thick, dashed, color = navyblue!20, forget plot] table [x index = 1, y index = 2, restrict y to domain= 0.8:1.0] {propagation_poi300.dat};

      \addplot [ultra thick,         color = darkred                 ] table [x index = 0, y index = 2                               ] {propagation_poi400.dat};
      \addplot [ultra thick, dashed, color = navyblue                ] table [x index = 1, y index = 2, restrict y to domain=-1  :0.8] {propagation_poi400.dat};
      \addplot [ultra thick, dashed, color = navyblue!20, forget plot] table [x index = 1, y index = 2, restrict y to domain= 0.8:1.0] {propagation_poi400.dat};

      \addplot [ultra thick,         color = darkred                 ] table [x index = 0, y index = 2                               ] {propagation_poi500.dat};
      \addplot [ultra thick, dashed, color = navyblue                ] table [x index = 1, y index = 2, restrict y to domain=-1  :0.8] {propagation_poi500.dat};
      \addplot [ultra thick, dashed, color = navyblue!20, forget plot] table [x index = 1, y index = 2, restrict y to domain= 0.8:1.0] {propagation_poi500.dat};

      \addlegendentry {simulation}
      \addlegendentry {stochastic approx.}
    \end {axis}
      \end {tikzpicture}}
    \caption {Erd\H os--R\'enyi network}
    \label {fig:evaluation_poi}
  \end {subfigure}
  \caption{Accuracy of the stochastic approximation under various parameter settings.
    Time unit is set so that $\mu = 1$.}
  \label{fig:evaluation}
\end {figure}
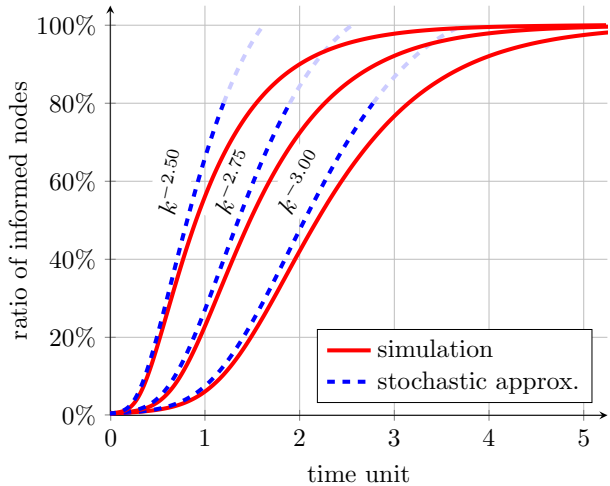
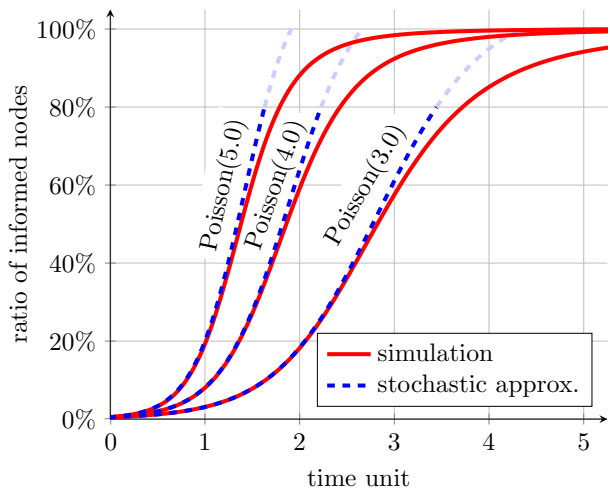

% comparisons to the simulation for scale-free and Erdos-Renyi networks
% accuracy: OK, but less for higher moments, explanation: Taylor
In order to evaluate the performance of the stochastic model, different degree distributions are visualized in Figure \ref {fig:evaluation} for scale-free and Erd\H os--R\'enyi networks.
The accuracy of the model is adequate in all cases, but as it was discussed in section \ref {sec_mathematicalmodel}, the first order Taylor-series approximation in \eqref {eq:krecvint} leads to greater and greater inaccuracies in the final phase.
Continuing the Taylor expansion could increase the accuracy, but the emergence of the covariance terms and higher moments would complicate the calculations.
In the final stage of the propagation, where the approximation is inaccurate, it is visualized with faded colors.

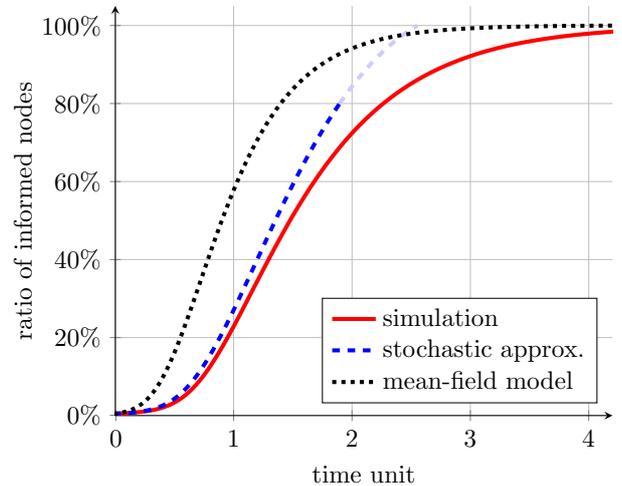
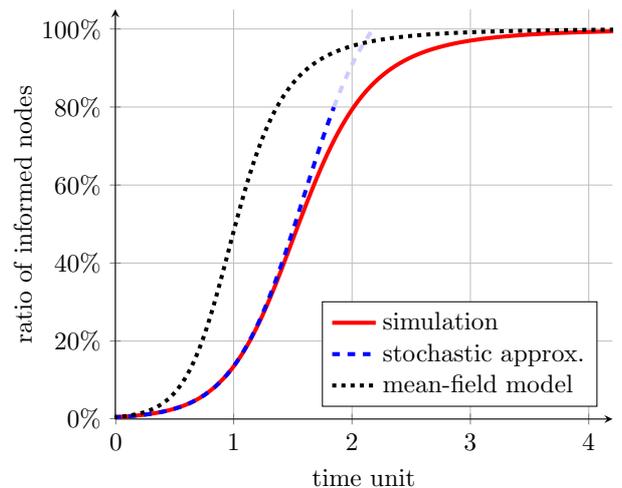
\begin {figure} [H] \centering
  \begin {subfigure} {0.45\textwidth} \centering
    \resizebox {\textwidth} {!} {
      \begin {tikzpicture}
        \begin {axis} [
          xmin=0, xmax=4.2, ymin=0, ymax=1.05,
          axis lines = middle,
          grid = both,
		   minor tick num=1,
          xlabel={time unit}, ylabel={ratio of informed nodes},
          xtick={0.005, 1, 2, 3, 4, 5},
          xticklabels={$0$, $1$, $2$, $3$, $4$, $5$},
          ytick={0.001, 0.2, 0.4, 0.6, 0.8, 1.0},
          yticklabels={$0\%$, $20\%$, $40\%$, $60\%$, $80\%$, $100\%$},
          xlabel style={anchor=north, at={(axis description cs:  0.50, -0.1)}},
          ylabel style={anchor=south, at={(axis description cs: -0.15,  0.5)}, rotate=90},
          legend pos=south east,
          legend cell align={left}
        ]
          \addplot [ultra thick,         color = darkred                 ] table [x index = 0, y index = 4                               ] {propagation_pow275.dat};
          \addplot [ultra thick, dashed, color = navyblue                ] table [x index = 1, y index = 4, restrict y to domain=-1  :0.8] {propagation_pow275.dat};
          \addplot [ultra thick, dashed, color = navyblue!20, forget plot] table [x index = 1, y index = 4, restrict y to domain= 0.8:1.0] {propagation_pow275.dat};
          \addplot [ultra thick, dotted, color = black                   ] table [x index = 2, y index = 4                               ] {propagation_pow275.dat};

          \addlegendentry {simulation}
          \addlegendentry {stochastic approx.}
          \addlegendentry {mean-field model}
        \end {axis}
      \end {tikzpicture}}
    \caption {Scale-free network: $P_{k_{\textrm {tot}}} \sim k^{-2.75}$}
    \label {fig:si_comparison_pow}
  \end {subfigure}
  \\[.5cm]
  \begin {subfigure} {0.45\textwidth} \centering
    \resizebox {\textwidth} {!} {
      \begin {tikzpicture}
        \begin {axis} [
          xmin=0, xmax=4.2, ymin=0, ymax=1.05,
          axis lines = middle,
          grid = both,
		   minor tick num=1,
          xlabel={time unit}, ylabel={ratio of informed nodes},
          xtick={0.005, 1, 2, 3, 4, 5},
          xticklabels={$0$, $1$, $2$, $3$, $4$, $5$},
          ytick={0.001, 0.2, 0.4, 0.6, 0.8, 1.0},
          yticklabels={$0\%$, $20\%$, $40\%$, $60\%$, $80\%$, $100\%$},
          xlabel style={anchor=north, at={(axis description cs:  0.50, -0.1)}},
          ylabel style={anchor=south, at={(axis description cs: -0.15,  0.5)}, rotate=90},
          legend pos=south east,
          legend cell align={left}
        ]
          \addplot [ultra thick,         color = darkred                 ] table [x index = 0, y index = 4                               ] {propagation_poi458.dat};
          \addplot [ultra thick, dashed, color = navyblue                ] table [x index = 1, y index = 4, restrict y to domain=-1  :0.8] {propagation_poi458.dat};
          \addplot [ultra thick, dashed, color = navyblue!20, forget plot] table [x index = 1, y index = 4, restrict y to domain= 0.8:1.0] {propagation_poi458.dat};
          \addplot [ultra thick, dotted, color = black                   ] table [x index = 2, y index = 4                               ] {propagation_poi458.dat};

          \addlegendentry {simulation}
          \addlegendentry {stochastic approx.}
          \addlegendentry {mean-field model}
        \end {axis}
      \end {tikzpicture}}
    \caption {Erd\H os--R\'enyi network: $P_{k_{\textrm {tot}}} \sim \mathrm {Poisson} (4.58)$}
    \label {fig:si_comparison_poi}
  \end {subfigure}
  \caption{Comparison of the results with that of the mean-field model.
    Time unit is set so that $\mu = 1$ in both cases.
    The parameters of the degree distributions were chosen so that the expected number of connections is the same in both networks.}
  \label{fig:si_comparison}
\end{figure}

% comparison to mean-field model
Let us have a look how the stochastic model and the degree-based SI mean-field model compare.
Figure \ref {fig:si_comparison} illustrates the predictions of the two models for a scale-free and an Erd\H os--R\'enyi network.
It can be seen, that the mean-field model underestimates the time needed for the information to get to the next node, especially at the beginning of the propagation.
The difference between the two models becomes even more significant when the higher moments of $K_{\textrm {ext}}$ are greater.
This is generally true if the degree distribution is wide and the propagation is at an early phase.
A low number of connections also leads to greater differences in accuracy as the connections used for the propagation -- the connections of the spanning tree connecting the informed nodes -- are not taken into account in the mean-field model.
The slopes of the curves match more accurately for the stochastic model, so it describes the mechanism of the propagation more precisely in this sense.

The major differences between the model presented here and the network-based SI model are the followings.
\begin {itemize}
\item {The SI model does not take into account the connections between the informed nodes responsible for the propagation.
Consequently, the SI model overestimates the number of connections on which the information spreads further.}
\item {Additionally, as a not informed node is more likely to be the next receiver node, if it has more connections with the informed node, the above difference is amplified further.}
\end {itemize}

\subsubsection* {Characteristics of the Propagation}

\begin {figure} [H] \centering
    \resizebox{0.45\textwidth}{!}{
      \begin {tikzpicture}
        \begin {axis} [
          xmin=0, xmax=4.2, ymin=0, ymax=1.05,
          axis lines = middle,
          grid = both,
		   minor tick num=1,
          xlabel={time unit}, ylabel={ratio of informed nodes},
          xtick={0.005, 1, 2, 3, 4, 5, 6, 7, 8},
          xticklabels={$0$, $1$, $2$, $3$, $4$, $5$, $6$, $7$, $8$},
          ytick={0.001, 0.2, 0.4, 0.6, 0.8, 1.0},
          yticklabels={$0\%$, $20\%$, $40\%$, $60\%$, $80\%$, $100\%$},
          xlabel style={anchor=north, at={(axis description cs:  0.50, -0.1)}},
          ylabel style={anchor=south, at={(axis description cs: -0.15,  0.5)}, rotate=90},
          legend pos=south east,
          legend cell align={left}
        ]
          \addplot [ultra thick, solid , color = darkred] table [x index = 0, y index = 4] {propagation_pow275.dat};
          \addplot [ultra thick, dashed, color = darkred] table [x index = 0, y index = 4] {propagation_poi458.dat};

          \addlegendentry {scale-free network}
          \addlegendentry {Erd\H os--R\'enyi network}
        \end {axis}
      \end {tikzpicture}} \caption {Information propagation in networks with scale-free ($P_{k_{\textrm {tot}}} \sim k^{-2.75}$) and Erd\H os--R\'enyi ($P_{k_{\textrm {tot}}} \sim \mathrm {Poisson} (4.58)$) networks based on simulation results. Time unit is set so that $\mu = 1$.} \label {poi_pow_comparison}
\end {figure}
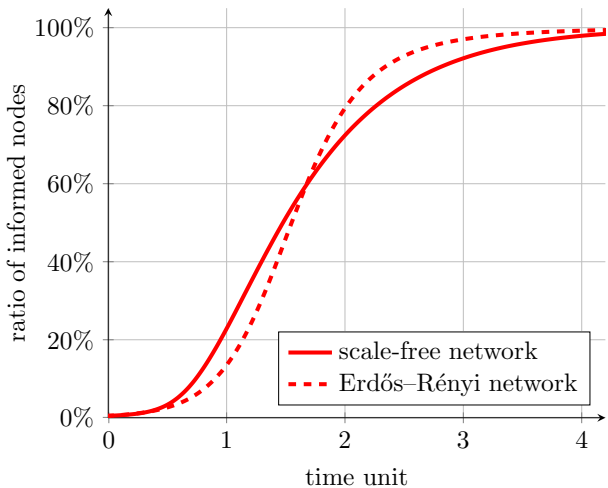

% comparison of power law and Poisson degree distribution
The comparison of scale-free and Erd\H os--R\'enyi networks is visualized in a Figure \ref {poi_pow_comparison}.
At the initial phase of the propagation high degree nodes tend to be informed, which increases the speed of the propagation.
In general, the expectation of the degree distribution of the informed nodes is the largest at the beginning, and it is greater than the expected value of the total degree distribution.
The propagation slows down as the average degree of the informed nodes declines due to the smaller degree nodes.
At the final stage the degree distribution gets closer to the total degree distribution.
It can be seen that -- with these parameters -- the propagation in the scale-free network cannot keep up with the propagation in the Erd\H os--R\'enyi network.
Although at the beginning the high degree nodes speeds up the propagation in the scale-free network, after a short initial phase the propagation declines as there is a significant number of nodes with small degrees.
In other words, the propagation reaches its final stage earlier in case of the Erd\H os--R\'enyi network.

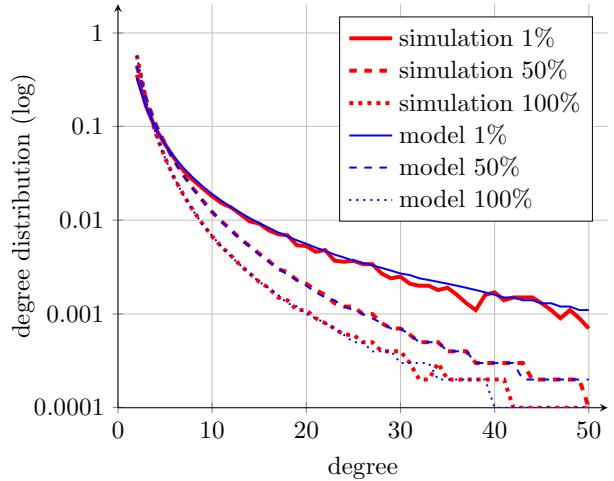
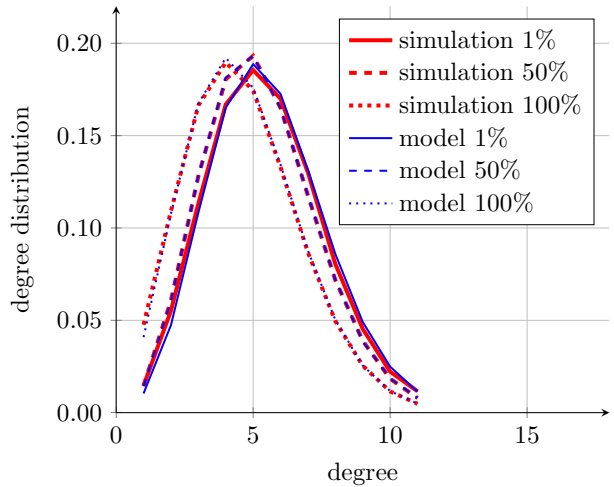
\begin {figure} [H] \centering
  \begin {subfigure} {0.45\textwidth} \centering
    \resizebox{\textwidth}{!}{
      \begin {tikzpicture}
        \begin {axis} [
          xmin=0, xmax=52, ymin=0.0001, ymax=2,
          ymode=log,
          axis y line = middle,
          axis x line = bottom,
          grid = both,
		   minor tick num=1,
          xlabel={degree}, ylabel={degree distribution (log)},
          xtick={0.005, 10, 20, 30, 40, 50},
          xticklabels={$0$, $10$, $20$, $30$, $40$, $50$},
          ytick={0.000101, 0.001, 0.01, 0.1, 0.99},
          yticklabels={$0.0001$, $0.001$, $0.01$, $0.1$, $1$},
          xlabel style={anchor=north, at={(axis description cs:  0.50, -0.1)}},
          ylabel style={anchor=south, at={(axis description cs: -0.15,  0.5)}, rotate=90},
          legend pos=north east,
          legend cell align={left}
        ]

          \addplot [ultra thick,         color = darkred ] table [x index = 0, y index =  1] {degree_distributions_pow275.dat};
          \addplot [ultra thick, dashed, color = darkred ] table [x index = 0, y index =  5] {degree_distributions_pow275.dat};
          \addplot [ultra thick, dotted, color = darkred ] table [x index = 0, y index =  9] {degree_distributions_pow275.dat};

          \addplot [thick,               color = navyblue] table [x index = 0, y index =  2] {degree_distributions_pow275.dat};
          \addplot [thick,       dashed, color = navyblue] table [x index = 0, y index =  6] {degree_distributions_pow275.dat};
          \addplot [thick,       dotted, color = navyblue] table [x index = 0, y index = 10] {degree_distributions_pow275.dat};

          \addlegendentry {simulation $1\%$}
          \addlegendentry {simulation $50\%$}
          \addlegendentry {simulation $100\%$}

          \addlegendentry {model $1\%$}
          \addlegendentry {model $50\%$}
          \addlegendentry {model $100\%$}
        \end {axis}
      \end {tikzpicture}}
    \caption {Scale-free network: $P_{k_{\textrm {tot}}} \sim k^{-2.75}$}
    \label {fig:dist_vs_time_scale_free}
  \end {subfigure}
  \\[.5cm]
  \begin {subfigure} {0.45\textwidth} \centering
    \resizebox{\textwidth}{!}{
      \begin {tikzpicture}
        \begin {axis} [
          xmin=0, xmax=18, ymin=0, ymax=0.22,
          axis lines = middle,
          grid = both,
		   minor tick num=1,
          xlabel={degree}, ylabel={degree distribution},
          xtick=      {0.005,  5 ,  10,   15 },
          xticklabels={$0$  , $5$, $10$, $15$},
          ytick=      {0.0001,  0.05 ,  0.10 ,  0.15 ,  0.20 },
          yticklabels={$0.00$, $0.05$, $0.10$, $0.15$, $0.20$},
          xlabel style={anchor=north, at={(axis description cs:  0.50, -0.1)}},
          ylabel style={anchor=south, at={(axis description cs: -0.15,  0.5)}, rotate=90},
          legend pos=north east,
          legend cell align={left}
        ]

          \addplot [ultra thick,         color = darkred ] table [x index = 0, y index =  1] {degree_distributions_poi458.dat};
          \addplot [ultra thick, dashed, color = darkred ] table [x index = 0, y index =  5] {degree_distributions_poi458.dat};
          \addplot [ultra thick, dotted, color = darkred ] table [x index = 0, y index =  9] {degree_distributions_poi458.dat};

          \addplot [thick,               color = navyblue] table [x index = 0, y index =  2] {degree_distributions_poi458.dat};
          \addplot [thick,       dashed, color = navyblue] table [x index = 0, y index =  6] {degree_distributions_poi458.dat};
          \addplot [thick,       dotted, color = navyblue] table [x index = 0, y index = 10] {degree_distributions_poi458.dat};

          \addlegendentry {simulation $1\%$}
          \addlegendentry {simulation $50\%$}
          \addlegendentry {simulation $100\%$}

          \addlegendentry {model $1\%$}
          \addlegendentry {model $50\%$}
          \addlegendentry {model $100\%$}
        \end {axis}
      \end {tikzpicture}}
    \caption {Erd\H os--R\'enyi network: $P_{k_{\textrm {tot}}} \sim \mathrm {Poisson} (4.58)$}
    \label {fig:dist_vs_time_erdos_renyi}
  \end {subfigure}
  \caption{Evolution of the degree distributions of the informed nodes.
      The degree distributions of the informed nodes given by the model and the Monte Carlo simulation are compared at three phases when $1\%$, $50\%$ and $100\%$ of the nodes are informed.}
  \label{fig:dist_vs_time}
\end {figure}

% evolution of the degree distribution
The analysis of the degree distributions of the informed nodes gives additional insight into the dynamics of the propagation.
Figure \ref {fig:dist_vs_time} shows how the degree distributions of informed nodes change in scale-free and Erd\H os--R\'enyi networks.
While the red curves show the experienced distributions from the simulation, the blue curves represent the calculated distributions.
As the three pairs of curves match well, the red curves are harder to spot.
In both cases, the degrees of the informed nodes tend to be greater at the beginning of the propagation since high-degree nodes are informed with higher probability.
As higher degree nodes tend to be informed earlier, the average degree among the informed nodes is always greater (or equal) to the average degree of the whole network.
As the propagation continues, the relative number of the lower-degree nodes increases, which shifts the degree distribution toward lower values.
Finally the information reaches all of the nodes; the degree distribution of the informed nodes becomes identical to the total degree distribution.
Due to the fat-tailed degree distribution of the scale-free network, there is a significant difference between the average degree of the informed nodes at the beginning of the propagation and the average degree of the total network.
As the nodes have similar degrees in the Erd\H os--R\'enyi networks, this difference is much smaller in this case.

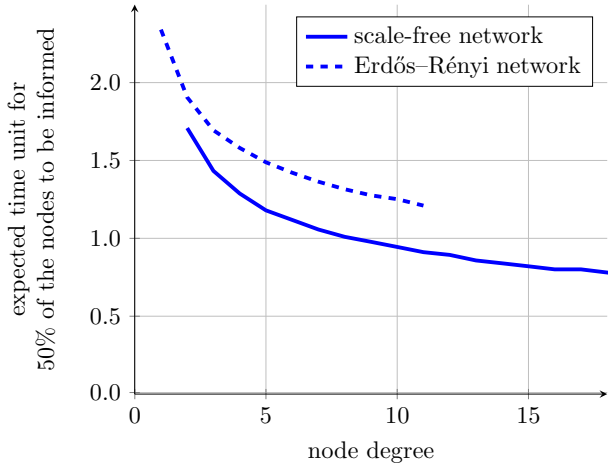
\begin {figure} [H] \centering
    \resizebox{0.45\textwidth}{!}{
      \begin {tikzpicture}
        \begin {axis} [
          xmin=0, xmax=18, ymin=0, ymax=2.5,
          axis lines = middle,
          grid = both,
		   minor tick num=1,
          xlabel={node degree}, ylabel={expected time unit for\\50\% of the nodes to be informed},
          xtick=      {0.005,  5 ,  10,   15 },
          xticklabels={$0$  , $5$, $10$, $15$},
          ytick=      { 0.01,  0.5 ,  1.0 ,  1.5,   2.0},
          yticklabels={$0.0$, $0.5$, $1.0$, $1.5$, $2.0$},
          xlabel style={anchor=north, at={(axis description cs:  0.50, -0.1)}},
          ylabel style={anchor=south, at={(axis description cs: -0.15,  0.5)}, align=center, rotate=90},
          legend pos=north east,
          legend cell align={left}
        ]
          \addplot [ultra thick,         color = navyblue] table [x index = 0, y index = 11] {degree_distributions_pow275.dat};
          \addplot [ultra thick, dashed, color = navyblue] table [x index = 0, y index = 11] {degree_distributions_poi458.dat};
          \addlegendentry {scale-free network}
          \addlegendentry {Erd\H os--R\'enyi network}
        \end {axis}
      \end {tikzpicture}} \caption {Average time after which $50\%$ of nodes are informed as a function of node degree in scale-free and Erd\H os--R\'enyi networks with degree distributions $P_{k_{\textrm {tot}}} \sim k^{-2.75}$ and $P_{k_{\textrm {tot}}} \sim \mathrm {Poisson} (4.58)$ respectively.} \label {median_times}
\end {figure}

% reception time vs node degree
It is also interesting to see how the degree of a node influences its reception time.
Figure \ref {median_times} shows when $50\%$ of the nodes with a given degree are informed.
In both of the networks, it is worth to increase the connection of a node in order to receive the information earlier.
The reception time highly depends on the number of degrees in case of small degree values.
If the degree of a node is high, an additional connection will not decrease the reception time significantly.

\section {Conclusions}

A new information propagation model was created which handles the information flow as independent Poisson processes on each connection.
Differential equations were derived describing the main characteristics of the information propagation, which were verified by Monte Carlo network simulations.
The comparison to the traditional degree-based mean-field model revealed the higher accuracy of the stochastic approach, which is more significant in sparse networks.
The results gave way to the analysis of information propagation in scale-free and Erd\H os--R\'enyi networks with the same number of connections in each.
It was shown that -- except for an initial period -- the information propagation is faster in the case of the Erd\H os--R\'enyi networks.
Finally, the evolution of the degree distribution of the informed nodes showed how the degree of a node influences its reception time.

\section* {Additional Information}

\paragraph {Author Contributions}
All work related to this paper was done by P\'eter L. Juh\'asz including conceptualization, methodology, software, validation, formal analysis, writing -- original draft, writing -- review \& editing and visualization.
The final manuscript was approved for publication.

\paragraph {Acknowledgment}
Special thanks to Dr. P\'eter B\'alint PhD, associate professor of Department of Stochastics, Budapest University of Technology and Economics for his contribution in deriving the number of external connections.

\paragraph {Competing Interests}
The author declares that he has no competing interests.

\paragraph {Funding}
This research did not receive any specific grant from funding agencies in the public, commercial, or not-for-profit sectors.

\end {multicols}

\pagebreak

\bibliographystyle {unsrt}

\end{document}